\documentclass[preprint,showpacs,preprintnumbers,amsmath,amssymb,natbib,showkeys]{revtex4}

\usepackage{graphicx}
\usepackage{epsfig}		
\usepackage{dcolumn}
\usepackage{bm}
\usepackage{subfigure}
\usepackage[english]{babel}
\def\0{\mbox{\tiny $0$}}
\def\1{\mbox{\tiny $1$}}
\def\2{\mbox{\tiny $2$}}
\def\3{\mbox{\tiny $3$}}
\def\4{\mbox{\tiny $4$}}
\def\5{\mbox{\tiny $5$}}
\def\6{\mbox{\tiny $6$}}
\def\7{\mbox{\tiny $7$}}
\def\8{\mbox{\tiny $8$}}
\def\9{\mbox{\tiny $9$}}

\def\f14{\mbox{\tiny $\frac{1}{4}$}}

\newcommand{\be}{\begin{equation}}
\newcommand{\ee}{\end{equation}}
\newcommand{\br}{\begin{eqnarray}}
\newcommand{\er}{\end{eqnarray}}

\newcommand{\bd}{\begin{displaymath}}
\newcommand{\ed}{\end{displaymath}}

\newcommand{\bfig}{\begin{figure}}
\newcommand{\efig}{\end{figure}}

\makeindex

\begin{document}

\title{Coherent quantum squeezing due to the phase space noncommutativity}

\author{A. E. Bernardini}
\email{alexeb@ufscar.br}
\author{S. S. Mizrahi}
\email{salomon@df.ufscar.br}

\affiliation{Departamento de F\'{\i}sica, Universidade Federal de S\~ao Carlos, PO Box
676, 13565-905, S\~ao Carlos, SP, Brasil}
\date{\today }

\begin{abstract}
The effect of phase space general noncommutativity on producing deformed coherent 
squeezed states is examined.
A two-dimensional noncommutative quantum system 
supported by a deformed mathematical structure similar to that of 
Hadamard billiards is obtained and their components behavior are 
monitored in time.
It is assumed that the independent degrees of freedom are 
two \emph{free} 1D harmonic oscillators (HO's), so the system Hamiltonian 
does not contain interaction terms. 
Through the noncommutative deformation parameterized by a Seiberg-Witten transform on the original canonical variables, 
one gets the standard commutation relations for the new ones, such that the obtained Hamiltonian 
represents then two \emph{interacting} 1D HO's. By assuming that one HO is inverted 
relatively to the other, it is shown that their effective interaction induces a 
squeezing dynamics for initial coherent states imaged in the phase space.
A suitable pattern of logarithmic spirals is obtained and some relevant properties 
are discussed in terms of Wigner functions, which are essential to put in 
evidence the effects of the noncommutativity.
\end{abstract}

\date{\today}
\maketitle

\section{Introduction}

Supported by a deformed Heisenberg-Weyl algebra \cite{Catarina,06A,07A,08A,09A,Gamboa,Guisado}, 
the phase space noncommutative generalization of quantum mechanics (QM) provides some 
elementary responses to typical issues which circumvent the intersection between quantum and 
classical mechanics. Besides evincing the role of noncommutativity in the predictions of 
the standard QM, some emblematic quantum effects like quantum decoherence, quantum 
entanglement \cite{Bernardini13B}, and the collapse of the wave function \cite{Bernardini13A} can be 
indeed fine-tuned to work as a probe of noncommutativity imprints on QM.

Even if it has been lately focused on studies of the quantum Hall effect \cite{Prange}, on the spectroscopy of the gravitational quantum well for ultra-cold neutrons \cite{06A}, on the Landau 
level and 2D harmonic oscillator problems in the phase space 
\cite{Nekrasov01,Rosenbaum,Bernardini13A}, and on quantumness and entanglement-separability 
issues \cite{Bernardini13B}, the noncommutativity is also believed to be a regular 
feature of quantum gravity and string theory \cite{Snyder47,Connes,Seiberg}.
Likewise, besides providing consistent explanations for the black hole singularity \cite{Bastos3} 
in the framework of quantum cosmology, the noncommutative QM scenario includes possible extensions 
of the matrix formulation of the uncertainty principle \cite{Ber2014}, and it has also 
stimulated a constructive analysis of the equivalence principle \cite{Bastos4}.
The framework is modeled on a $2n$-dimensional phase space where the time variable is assumed as a commutative parameter, 
and the phase space coordinate commutation relations are supported by a noncommutative 
algebra, in manner that a noncommutative formulation of QM is more suitably stablished in terms 
of the Weyl-Wigner-Groenewald-Moyal (WWGM) formalism \cite{Groenewold,Moyal,Wigner}.

In this contribution, the role of the noncommutative algebra of two free harmonic 
oscillators (HO's) (described by a Hamiltonian with a quadratic structure involving 
the phase space variables, positions and momenta) on producing squeezing is discussed through an analysis based on a time-evolving Wigner function. 
A Seiberg-Witten transform on the noncommutative variables \cite{Seiberg} leads a novel set of (now canonical) variables which exhibit the standard commutation relations of Weyl-Heisenberg algebra, at the price that now the HO's are not more free, and they interact through the emergence of an additional term in the Hamiltonian. 
Thus, this procedure allows one to determine how the noncommutative parameters induce the squeezing 
dynamics for initial coherent states by the arising of a specific interaction. 
The other way around, one could say that two interacting HO's in QM are equivalent 
to two free ones whose phase space variables follow a generalized noncommutative algebra. 
Last but not least, it is worth reminding that the system can be circumstantially 
identified with a Hadamard dynamical system \cite{Hadamard}. 

\section{The noncommutative algebra of a dynamical system}

The Hamiltonian formulation of 2D quantum mechanical problems correspond to the most 
accessible systems for which the noncommutative phase space properties can be probed 
\cite{Nekrasov01,Rosenbaum,Bernardini13A}. Therefore, one considers 
two 1D HO's sliding frictionlessly, and having the Hamiltonian,
\begin{equation}
{H}(\mathbf{q},\mathbf{p}) = \sum_{i,j=1}^{2}g^{ij}\left(\frac{1}{2m}\, p_i p_j + 
\frac{1}{2}m \omega^2\, q_i q_j \right),
\end{equation}
where the operator vector notation is set as $\mathbf{v} = ({v}_1,{v}_2)$, and $g^{ij}$ 
is the metric tensor on the manifold.  
Through a particular choice of the Riemann manifold parameterized by $g^{ij}$, the above 
Hamiltonian can be converted into suitable probe of noncommutative effects. One thus sets 
$g^{ij} = \delta^{i1}\delta^{j2} + \delta^{i2}\delta^{j1}$ such that it
shall then represent a pair of $45$ degrees rotated decoupled HO's, one with 
corresponding energy spectrum unbounded from below, and another with energy spectrum 
unbounded from above. This problem was treated in a different context by 
R. J. Glauber in \cite{Glauber}. In fact, by identifying the $1D$ harmonic oscillator 
Hamiltonian with
\begin{equation}
{H}_{HO}(x_j,k_j) =\frac{1}{2m}\, k_j^{\2} + \frac{1}{2}m \omega^2\,  x_j^{\2},
\end{equation}
with $x_j = (q_1 - (-1)^{j}q_2)/\sqrt{2}$ and $k_j = 
(p_1 - (-1)^{j}p_2)/\sqrt{2}$, with $j = 1,\,2$, one has
\begin{equation}
{H}(\mathbf{q},\mathbf{p}) \equiv  {H}_{HO}(x_1,k_1) - {H}_{HO}(x_2,k_2)
\equiv {H}_{HO}(x_1,k_1) + {H}_{HO}(i\,x_2,i\,k_2),
\end{equation}
where the last passage indicates that the system labeled by $j = 2$ can be read as 
a Hamiltonian component that is presumed to have its position and momentum coordinates driven 
by a Wick rotation, which turns a bounded Hamiltonian into an unbounded one, 
from below. Globally, it corresponds to change a spherical manifold, namely the 
simplest compact Riemann surface with positive curvature, into a hyperbolic 
manifold, by the way, a compact Riemann surface with negative curvature.

One shall notice that the noncommutative deformation induces some modifications that 
allow one to overcome the infinities and divergent behaviors originated from 
the above Hamiltonian dynamics. The spatial and momentum noncommutative 
algebra is set as
\begin{equation}
\left[{q}_i,  {q}_j \right] = i \theta \epsilon_{ij}\hspace{0.2 cm}, \hspace{0.2 cm} 
\left[  {q}_i,  {p}_j \right] = i \delta_{ij}\hbar , \hspace{0.2 cm} 
\left[  {p}_i,  {p}_j \right] = i \eta \epsilon_{ij}\hspace{0.2 cm}, 
\hspace{0.2 cm} i,j=1,2,
\label{aebnnnccc}
\end{equation}
with the Levi-Civita tensor $\epsilon_{ij} = -\epsilon_{ji}$, such that the 
Seiberg-Witten (SW) \cite{Seiberg} map to the commutative operators, 
$\{\mbox{\bf \em Q},\mathbf{\Pi}\}$, can be read as
\begin{equation}
{q}_i = \lambda  \mathit{{Q}}_i - \frac{\theta}{2\lambda \hbar} \sum_{j=1}^{2}\epsilon_{ij} {\Pi}_j 
\hspace{0.5 cm},\hspace{0.5 cm}  {p}_i = \mu {\Pi}_i + \frac{\eta}{2 \mu \hbar} 
\sum_{j=1}^{2}\epsilon_{ij}  \mathit{{Q}}_j~,
\label{aebSWmap}
\end{equation}
which is invertible when a constraint on the dimensionless parameters $\lambda$ and $\mu$ is 
stablished by the relation \cite{Catarina}
\begin{equation}
\frac{\theta \eta}{4 \hbar^2} = \lambda \mu ( 1 - \lambda \mu ),
\label{aebconstraint}
\end{equation}
as to have the corresponding Jacobian given by 
\begin{equation}
\left\| \frac{\partial ({q},{p})}{\partial 
({\mathit{Q}}, {\Pi})} \right\|= (\det {\mathbf \Omega})^{1/2}= 1 - 
\frac{\theta \eta}{\hbar^2},
\label{aebconstraint2}
\end{equation}
with
$$\mathbf \Omega= \left[
\begin{array}{cccc}
0&+\theta/\hbar&+1&0\\
-\theta/\hbar&0&0&+1\\
-1&0&0&+\eta/\hbar\\
0&-1&-\eta/\hbar&0
\end{array}
\right],$$
where $0 \leq \theta\eta < \hbar^2$.
One thus obtains the inverse map given by \cite{Catarina}
\begin{eqnarray}
\mathit{{Q}}_i &=& \mu \left(1 - \frac{\theta \eta}{\hbar^2} \right)^{- 1 / 2} 
\left( {q}_i + \frac{\theta}{2 \lambda \mu \hbar} \sum_{j=1}^{2}\epsilon_{ij}  {p}_j \right),\nonumber\\
{\Pi}_i &=& \lambda \left(1 - \frac{\theta \eta}{\hbar^2} \right)^{-1 / 2} 
\left( {p}_i-\frac{\eta}{2 \lambda \mu \hbar} \sum_{j=1}^{2}\epsilon_{ij}  {q}_j \right),
\label{aebSWinverse}
\end{eqnarray}
which guarantees that the new coordinates satisfy the standard Weyl-Heisenberg algebra,
\begin{equation}
\left[{Q}_i,  {Q}_j \right] =  \left[  {\Pi}_i,  {\Pi}_j \right] = 0 \qquad \mbox{and} \qquad
 \left[  {Q}_i,  {\Pi}_j \right] = i \delta_{ij}\hbar, \hspace{0.2 cm} i,j=1,2,
\end{equation}
such that the Hamiltonian of the previously uncoupled HO's can be 
re-written in terms of the new variables, $\mathit{{Q}}_i$ and ${\Pi}_i$, as 
\begin{equation}
H(\mbox{\bf \em Q},\mathbf{\Pi}) = \sum_{i,j=1}^{2} g^{ij}\,\left(\alpha^2 \mathit{{Q}}_i \mathit{{Q}}_j + \beta^2 {\Pi}_i {\Pi}_j \right) + \frac{\Gamma}{2} 
\left( \left\{\mathit{{Q}}_1,\ {\Pi}_1 \right\}
-\left\{\mathit{{Q}}_2,\ {\Pi}_2 \right\} \right),
\label{aebHamilton}
\end{equation}
with \footnote{Notice the {\em minus} sign that replaces the {\em plus} sign 
of the corresponding results for the 2D harmonic oscillator from \cite{Bernardini13A}.}
\begin{eqnarray}
{\alpha}^2 &\equiv& \frac{\lambda^2 m \omega^2}{ 2} - \frac{\eta^2}{ 8m \mu^2 \hbar^2}~,\label{aebalfa2}\\
{\beta}^2 &\equiv& \frac{\mu^2}{ 2m} - \frac{m \omega^2 \theta^2}{8 \lambda^2 \hbar^2}~,\label{aebbeta2}
\end{eqnarray}
where, preliminarily assuming that the constraint (\ref{aebconstraint}) is satisfied, the positiveness of ${\alpha}^2$ and ${\beta}^2$ are independently and
phenomenologically assumed \emph{ad hoc}, with the consequences of (\ref{aebconstraint}) straightforwardly extended to the choice of the parameters $\lambda$ and $\mu$, and with 
\begin{equation}
{\Gamma} \equiv \frac{\theta}{2\hbar}m \omega^2  - \frac{\eta}{2m\hbar},
\label{aebparam}
\end{equation}
the parameter that couples the HO's.
The Hamiltonian remains unbounded in the noncommutative scenario, however, the noncommutative algebra from Eq.(\ref{aebnnnccc})induces an additional coupling between the unbounded subsystems mediated by the parameter $\Gamma$, as to have an isentropic system with globally conserved energy flows recursively absorbed from the one to the other, as it shall be depicted from the solutions of the equations of motion.

Since $\mbox{\bf \em Q}$ and 
$\mathbf{\Pi}$ satisfy Hamilton equations of motion, 
one has the following set of coupled first-order differential equations,
\begin{eqnarray}
\dot{\Pi}_k &=& -\frac{i}{\hbar} \left[\Pi_k,\,H\right] = ~~ (-1)^{k}\, 
\left(2 \alpha^2\,\sum_{j=1}^{2}\epsilon_{kj}\mathit{Q}_j + \Gamma\,\Pi_k\right),\nonumber\\
\dot{\mathit{Q}}_k &=& -\frac{i}{\hbar} \left[\mathit{Q}_k,\,H\right] = 
-(-1)^{k}\, \left(2 \beta^2\,\sum_{j=1}^{2}\epsilon_{kj}\Pi_j + \Gamma\,\mathit{Q}_k\right),
\qquad k = 1,2. \label{aebeqs01}
\end{eqnarray}
After simple mathematical manipulations, the above equations can be written 
as two uncoupled second-order differential equations,
\begin{eqnarray}
\ddot{\Pi}_k -  (-1)^{k}\, 2 \Gamma\,\dot{\Pi}_k + (\Gamma^2 + 
4\alpha^2\beta^2)\,\Pi_k  &=& 0, \nonumber\\
\ddot{\mathit{Q}}_k + (-1)^{k}\, 2 \Gamma\,\dot{\mathit{Q}}_k + (\Gamma^2 + 
4\alpha^2\beta^2)\,\mathit{Q}_k &=& 0,
\label{aebeqs02}
\end{eqnarray}
from which one gets the dynamical variables,
\small
\begin{subequations} \label{aebsub}
\begin{eqnarray}
\mathit{Q}_1(t)&=& \exp{(+\Gamma t)} \left[x\,\cos(\Omega t)  + 
\frac{\beta}{\alpha}\pi_y\,\sin(\Omega t)\right],\label{aebsub1}\\
\mathit{Q}_2(t)&=& \exp{(-\Gamma t)} \left[y\,\cos(\Omega t)  + 
\frac{\beta}{\alpha}\pi_x\,\sin(\Omega t)\right],\label{aebsub2}\\
\Pi_1(t)&=& \exp{(-\Gamma t)} \left[\pi_x\,\cos(\Omega t)  - 
\frac{\alpha}{\beta}y\,\sin(\Omega t)\right],\label{aebsub3}\\
\Pi_2(t)&=& \exp{(+\Gamma t)} \left[\pi_y\,\cos(\Omega t)  - 
\frac{\alpha}{\beta}x\,\sin(\Omega t)\right],
\label{aebsub4}
\end{eqnarray}
\end{subequations}
\normalsize
where $x = Q_1(0),\, y= Q_2(0),\,\pi_x = \Pi_1(0),$ and $\pi_y = \Pi_2(0)$ 
are the initial conditions, and 
\begin{equation}
\Omega   = 2 \alpha \beta  =  \omega \sqrt{(2\lambda\mu - 1)^2 - \varepsilon^2},
\label{aebeq37}
\end{equation}
with
\begin{equation}
\varepsilon = \frac{1}{2\hbar}\left[m\omega\theta - \frac{\eta}{m\omega}\right],
\end{equation}
so that $\Gamma = \omega \varepsilon$. Notice that the mathematical structure of the above results is very similar to that from Ref.~\cite{Bernardini13A}, for which a $2D$ noncommutative HO is discussed. One observes that, for $\varepsilon > 0$, one variable (for each HO) is amplified as time goes on whereas the other is attenuated, such 
that the commutation relations remain unaffected, 
$\left[Q_i(t),\Pi_j(t)\right]= i \hbar \delta_{ij}$. By setting $\varepsilon = 0$ 
one recovers the solutions for the uncoupled HO's coordinates. For 
$0 < \varepsilon \ll 1$, one has $$\Omega \sim 
\omega[1 + \mathcal{O}(\varepsilon^2)]\times\vert2\lambda\mu - 1\vert \sim 
\omega[1 + \mathcal{O}(\theta^2,\, \eta^2,\, \theta\eta)],$$ and the noncommutative parameters, 
$\theta$ and $\eta$, introduce second-order corrections in $\Omega$ (c. f. Ref.~\cite{Bernardini13A}). 
Likewise, the modifications due to $\Gamma = \omega \varepsilon$ correspond 
to typical first order effects as quantified in Refs.~\cite{Catarina, Bernardini13A}.

\section{Phase space and Wigner function}

\begin{figure}
\includegraphics[width= 7.cm]{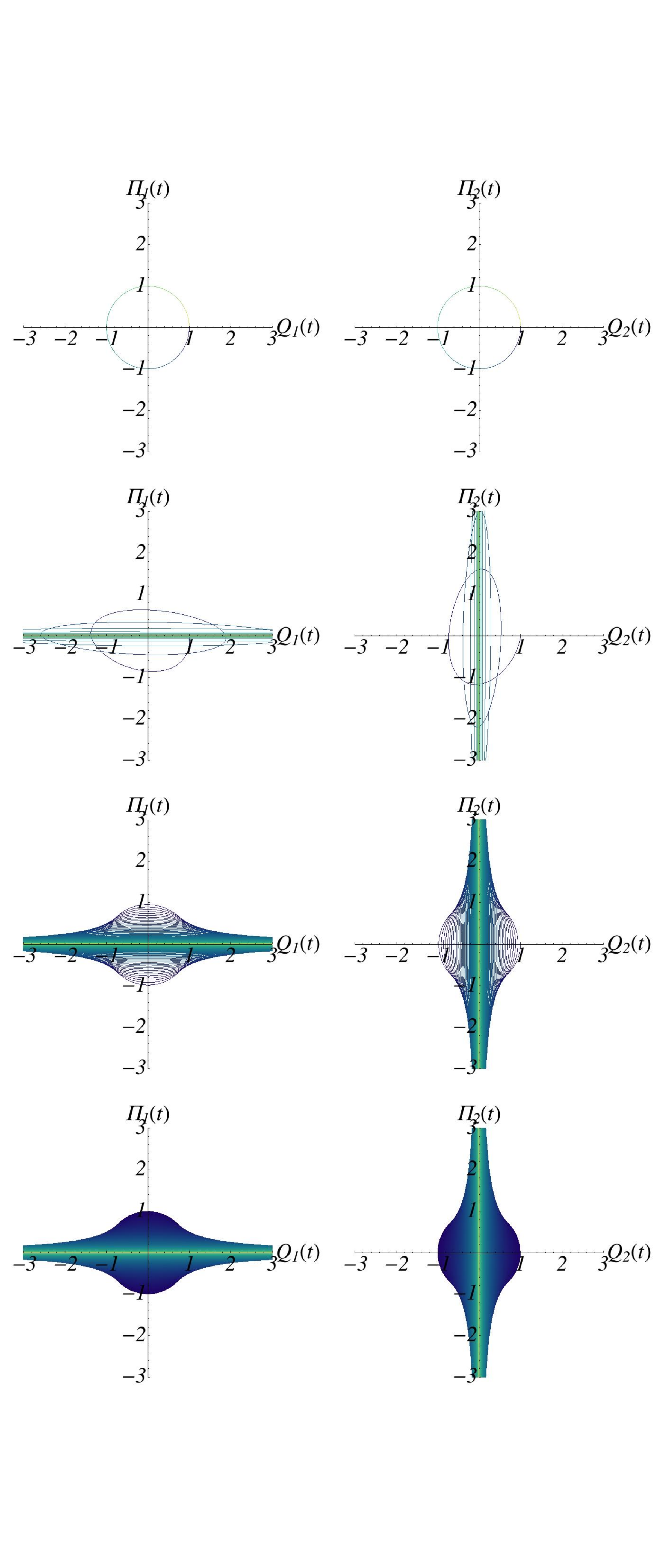}
\hspace{.01 cm}
\includegraphics[width= 7.cm]{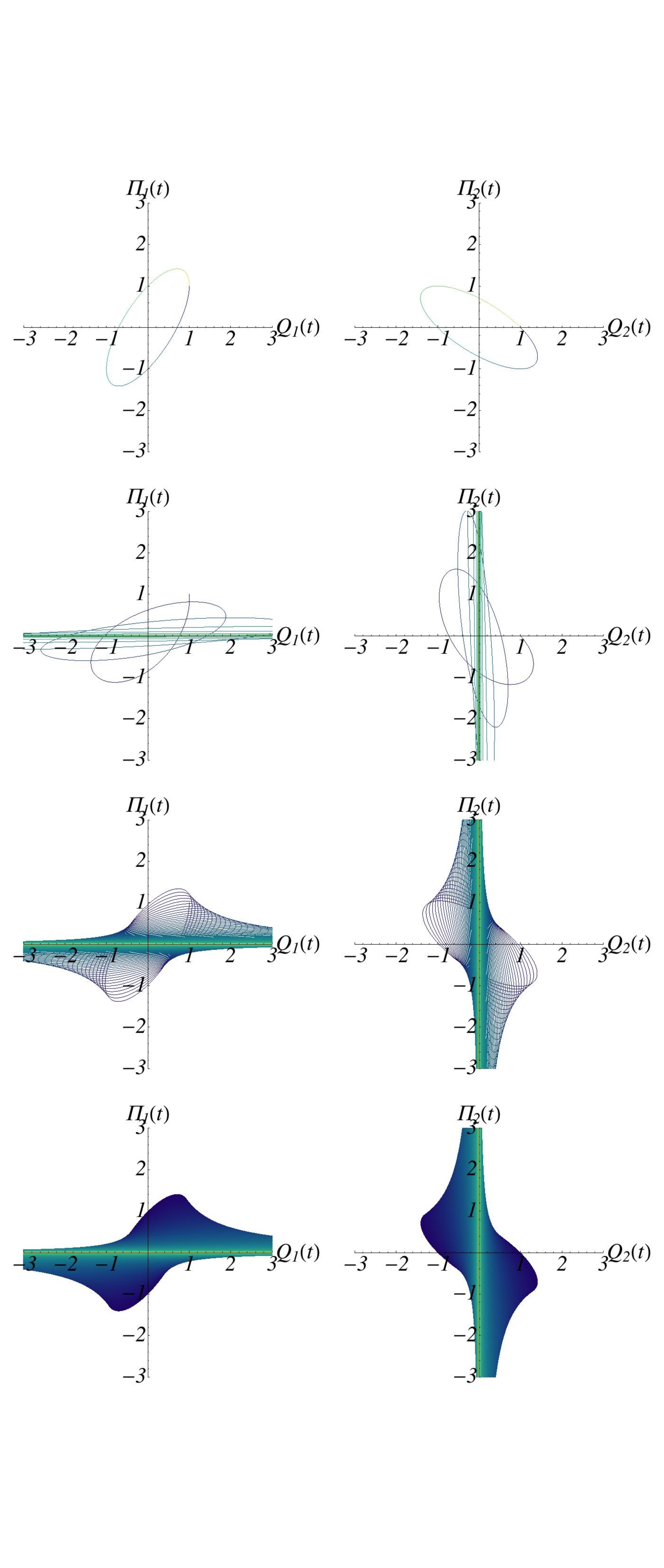}
\includegraphics[width= 1.5cm]{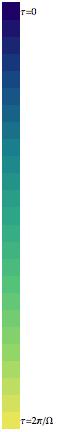}
\vspace{1.0 cm}
\caption{\footnotesize  (Color online) Time evolution of the phase space 
coordinates,$(\mathit{Q}_1(t),\Pi_1(t))$ and $(\mathit{Q}_2(t),\Pi_2(t))$.
The plots in the first line of each set (column) refer to the phase space 
elliptical orbits similar to those described for 2D harmonic oscillators 
(as if one had set $\epsilon = 0$ in the noncommutative map). From the second to the 
forth plot lines one has set arbitrary values for $\epsilon$, $\epsilon = 
1/10,\, 1/100$, and $1/1000$ respectively. Positive and time reversed 
logarithmic spirals describe the time-evolving open orbits for these cases.
One has used a {\em BlueGreenYellow} ({\em GrayLevel}) scale in order to 
denote the time scale, $\tau$, varying from $0$ (blue (dark gray)) to 
$2 \pi / \Omega$ (yellow (light gray)), such that orbits start and 
finish at $(x,\pi_x, y,\pi_y)$ equals to
$(1,0,1,0)$ (first column) and $(1,1,1,0)$ (second column).
By convenience, one has set $\alpha = \beta$, that is equivalent to $m\omega = \hbar = 1$.}
\label{aebPhase}
\end{figure}

The time evolution within the phase space associated with the operators  
$(\mathit{Q}_1(t),\Pi_1(t))$ and $(\mathit{Q}_2(t),\Pi_2(t))$ are depicted 
in Fig.~\ref{aebPhase}, for which the time is in the range $[0, 2\pi/\Omega]$. 
For convenience, the auxiliary variable
\begin{equation}
\epsilon = \frac{\Gamma}{\Omega} = \frac{\varepsilon}{\sqrt{(2\lambda\mu - 1)^2 - 
\varepsilon^2}},
\end{equation}
is defined to be used for a non-perturbative analysis of the results. The phase space maps 
from the first and second columns in Fig.~\ref{aebPhase} correspond, respectively, 
to direct and indirect logarithmic spirals which are associated to damping and 
amplifying modes. Two examples for which one identifies different choices of 
the set of initial conditions are presented. Considering only one  
separated HO, one gets it as an open (unbounded from below Hamiltonian, or even 
non-Hamiltonian) system. To reestablish the canonical formalism and 
conservation of information, one must have both HO's in order to have a closed 
and isentropic system.

The dynamical evolution of a wavefunction or a density operator can be mapped into a 
Wigner function (WF), $W(\mbox{\bf \em Q},\mathbf{\Pi})$ (now on the variables $\bf{Q}$ 
and $\bf{P}$ are c-numbers), since one can follow trajectories of the motion 
in the phase space. 
One has only to ensure that each point of the WF moves in the correlated 
paths, $1 \leftrightarrow 2$, as depicted for instance in the plots from Fig.~\ref{aebPhase}. 
This is reflected by a characteristic invariance property of stationary 
WFs.
The time evolution of a WF is given by a propagator acting on an ``initial'' one,
\be
W(\mathbf{Q},\mathbf{\Pi},t) = e^{-i\mathcal{L}_Q (t-t_0)}W(\mathbf{Q},\mathbf{\Pi},t_0),
\ee
where
\be
\mathcal{L}_Q \equiv H\left(\mathbf{Q},\mathbf{\Pi} \right) 
\left[ i\frac{2}{\hbar} \sin \left(\frac{\hbar}{2} \overleftrightarrow{\Lambda }\right) \right],
\ee
is the Liouvillian superoperator, $H\left(\mathbf{Q},\mathbf{\Pi} \right)$ 
is Weyl's map of the Hamiltonian operator, and
\be
\overleftrightarrow{\Lambda } = \frac{\overleftarrow{\partial}}{\partial{\mathbf{Q}}}
\cdot \frac{\overrightarrow{\partial}}{\partial{\mathbf{\Pi}}}
- \frac{\overleftarrow{\partial}}{\partial{\mathbf{\Pi}}}
\cdot \frac{\overrightarrow{\partial}}{\partial{\mathbf{Q}}},
\ee
is an operator acting on the left on $H\left(\mathbf{Q},\mathbf{\Pi} \right)$ 
and on the right on the WF. For a quadratic Hamiltonian, the Liouvillian reduces to
\be
\mathcal{L}_{cl} = i H\left(\mathbf{Q},\mathbf{\Pi} \right) 
\overleftrightarrow{\Lambda },
\ee
resulting in a classical evolution (this is another form of the Ehrenfest theorem)
\cite{BMT1,BMT2}.
Thus, assuming $t_0=0$,
\br
W(\mathbf{Q},\mathbf{\Pi},t) &=& e^{-i\mathcal{L}_{cl} (t)}W(\mathbf{Q},\mathbf{\Pi},0) \nonumber \\
&=& W(e^{i\mathcal{L}_{cl} (t)}\mathbf{Q}(0),\ e^{i\mathcal{L}_{cl} (t)}\mathbf{\Pi}(0),0) 
= W(\mathbf{Q}(-t),\mathbf{\Pi}(-t),0),
\er
using the time reversed solutions of Eqs.~(\ref{aebsub}), with the q-numbers being substituted 
by c-numbers, and reminding that 
$\left\{ \mathit{Q}_1(0), \mathit{\Pi}_1(0), \mathit{Q}_2(0), \mathit{\Pi}_2(0) \right\}
\equiv \left\{ \mathit{x, \pi_x, y, \pi_y} \right\}$.

\begin{figure}
\vspace{-1.0 cm}
\includegraphics[width= 6. cm]{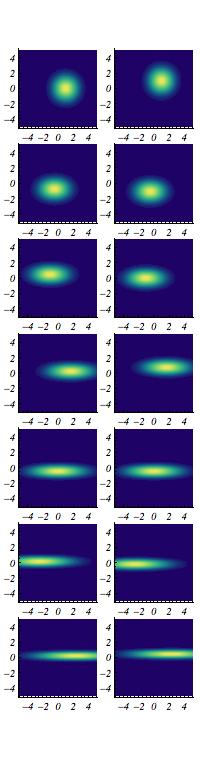}
\vspace{-2.0 cm}
\caption{\small  (Color online) Coherent quantum squeezing for Gaussian states 
in the $\mathit{Q}_1-\Pi_1$ plane, which evolves in correspondence with the phase 
space map depicted in Fig.~\ref{aebPhase}. At time $\tau = 0$ the Wigner function 
is assumed to be centered at departing points $(x,\pi_x)$ from Fig.~\ref{aebPhase}. 
One has considered time intervals such that $\tau = k \pi(32\epsilon \Omega)^{-1}$, 
with $k$ from $0$ to $6$, and $\epsilon = 1/10$, in order to reproduce an equally 
spaced time evolution sequence of plots. The contour plot follows a 
{\em BlueGreenYellow} scale (from yellow (light gray) which corresponds 
to $1$, to blue (dark gray) which corresponds to $0$), from which the 
squeezing effect can be easily noticed.} \label{aebMapa01}
\end{figure}

It has been attributed to $W(\mathbf{Q},\mathbf{\Pi},0)$ a symmetric (non-squeezed) Gaussian form 
for both the HO's, and one looks over to the time evolution of HO 1 only, getting the marginal 
Wigner functions $\tilde{W}^{(1)}(\mathit{Q}_1,\mathit{\Pi}_1;t)$ 
as shown in Fig.~\ref{aebMapa01}.
As time goes on 
the WF evolves to a {\em squeezed} state,   
and the squeezing dynamics follows the logarithmic spiral 
evolution of the phase space variables (c. f. Fig.~\ref{aebPhase}), and 
the results do not depend quantitatively on the parameter $\epsilon$ 
since one has chosen scale independent values for $\tau$.

\section{Summary and discussions}

The implications of noncommutativity of the dynamical variables of two \emph{free} HO's 
on producing squeezing states has been established. This feature is observed 
when a Seiberg-Witten transform is applied on the variables, thus \emph{coupling} the HO's. 
The squeezing observed in each HO occurs by a phase difference of $\pi$, due to the 
different exponential factor in Eqs.~(\ref{aebsub}). 

Looking at only one HO, the missing information is completely absorbed 
by the other, once it flows recursively from the one to the other although it is 
globally conserved, since the system is isentropic, as it was already discussed, 
in a similar context, for two interacting modes of the electromagnetic field, 
in Refs. \cite{OMD,CDM}. 

Our results also reinforce previous analysis where the noncommutative effects 
of variables can be considered when addressing the issues of the fine tuning 
of quantum effects.
Squeezing and quantum dissipation properties \cite{Cel01,Cel02} and a set of analogous results outside the scope of the noncommutative can also be found in the study of a 
$SU(1,1)$ structure of squeezed states as damped oscillators \cite{Cel03} dynamically generated by single-mode hamiltonians characterized by two-photon process interactions, with damping elements similar to that exhibited by Eq.~(\ref{aebHamilton}).
Recently, the self-similarity properties of fractals 
has been discussed in the context of the theory of entire analytical 
functions and of deformed algebra of coherent states \cite{Vitiello}, and their 
functional realization in terms of squeezed coherent states has been obtained. 
The noncommutativity in the phase space reported in this 
paper changes the similar dynamics of two uncoupled 1D HO's
(Hadamard's billiard) into the dynamics of coupled logarithmic spirals.

The theoretical description supported by the noncommutativity in the 
phase space then supports a consistent explanation for several experimental 
observations of temporal large scale effects in superconductors, crystals, 
ferromagnets, etc \cite{Vitiello4}, where squeezed states also appear.
As expected, the relevance of these results in terms of their experimental feasibility/detectability may depend essentially on the noncommutative parameters $\theta$ and $\eta$.

\begin{acknowledgments}
The work of AEB is supported by the Brazilian Agency CNPq under the grants 300809/2013-1 and 440446/2014-7. The work of SSM is supported by CNPq and FAPESP, Brazilian agencies, through the INCT program.
\end{acknowledgments}

\pagebreak
\newpage

\renewcommand{\baselinestretch}{1}

\begin{thebibliography}{99}

\bibitem{Catarina}
C. Bastos, O. Bertolami, N. C. Dias and J. N. Prata, J. Math. Phys. {\bf 49}, 072101 (2008). 

\bibitem{06A}
O. Bertolami, J. G. Rosa, C. M. L. de Arag\~ao, P. Castorina and D. 
Zappal\`a, Phys. Rev. {\bf D72}, 025010 (2005).

\bibitem{07A}
O. Bertolami, J. G. Rosa, C. Arag\~ao, P. Castorina and D. Zappal\`a,
Mod. Phys. Lett. {\bf A 21}, 795 (2006).

\bibitem{08A}
N. C. Dias and J. N. Prata, Annals Phys. {\bf 324}, 73 (2009).

\bibitem{09A}
C. Bastos and O. Bertolami, Phys. Lett. {\bf A372}, 5556 (2008).

\bibitem{Gamboa} 
J. Gamboa, M. Loewe and J. C. Rojas, Phys. Rev. {\bf D64}, 067901 (2001);
J. Gamboa {\it et al.}, Mod. Phys. Lett. {\bf A16}, 2075 (2001).

\bibitem{Guisado}
O. Bertolami, L. Guisado, JHEP {\bf 0312}, 013 (2003).

\bibitem{Bernardini13B}
C. Bastos, A. E. Bernardini, O. Bertolami, N. C. Dias and J. N. Prata, Phys. 
Rev. {\bf D88}, 085013 (2013).

\bibitem{Bernardini13A}
A. E. Bernardini and O. Bertolami, Phys. Rev. {\bf A 88} 012101, (2013).

\bibitem{Prange}
R. Prange and S. Girvin, {\em The Quantum Hall Effect}, (Springer, New York, 1987).

\bibitem{Nekrasov01}
M. R. Douglas and N. A. Nekrasov, Rev. Mod. Phys. {\bf 73}, 977 (2001).

\bibitem{Rosenbaum}
M. Rosenbaum, J. David Vergara and L. Roman Juarez, Phys. Lett. {\bf A367}, 1 (2007);
M. Rosenbaum and J. David Vergara, Gen. Rel. Grav. {\bf 38}, 607 (2006).

\bibitem{Snyder47}
H. S. Snyder, Phys. Rev. {\bf 71}, 38 (1947).

\bibitem{Connes}
A. Connes, M. R. Douglas and A. Schwarz, JHEP {\bf 02}, 003 (1998);
M. R. Douglas and C. Hull, JHEP {\bf 02}, 008 (1998);
V. Schomerus, JHEP {\bf 9906}, 030 (1999).

\bibitem{Seiberg} 
N. Seiberg and E. Witten, JHEP {\bf 9909}, 032 (1999).

\bibitem{Bastos3}
C.Bastos, O. Bertolami, N. C. Dias and J. Prata, Phys. Rev. {\bf D78}, 023516 (2008);
Phys. Rev. {\bf D80}, 124038 (2009); Phys. Rev. {\bf D82}, 041502 (2010);
Phys. Rev. {\bf D84}, 024005 (2011).

\bibitem{Ber2014}
C. Bastos, A. E. Bernardini, O. Bertolami, N. C. Dias and 
J. N. Prata, Phys. Rev.{\bf D90}, 045023 (2014). 

\bibitem{Bastos4}
C. Bastos, O. Bertolami, N. C. Dias and J. Prata, Class. Quant. Grav. {\bf 28}, 125007 (2011).

\bibitem{Groenewold} 
H. Groenewold, Physica {\bf 12} (1946) 405.

\bibitem{Moyal} 
J. E. Moyal, Proc. Camb. Phil. Soc. {\bf 45} (1949) 99.

\bibitem{Wigner}
E. Wigner, Phys. Rev. {\bf 40} (1932) 749.

\bibitem{Hadamard}
J. Hadamard, {\em Les surfaces \`{a} courbures oppos\'{e}es et leurs lignes 
g\'{e}od\'{e}siques}, J. Math. Pures et Appl. {\bf 4}, 27 (1898).

\bibitem{Glauber} R. J. Glauber, \emph{Ampifiers, Attenuators, and Schrödinger´s Cat} 
in ``New Techniques and Ideas in Quantum Measurement Theory´´, Annals of the New York 
Academy of Science \textbf{480}, 336 (1986).

\bibitem{BMT1} G. Bund, S. S. Mizrahi, and M. C. Tijero, \emph{Phys. Rev. A} {\bf 53}, 
1191 (1996).

\bibitem{BMT2} G. Bund, S. S. Mizrahi, and M. C. Tijero, \emph{Phys. Rev. A} {\bf 56}, 
2825 (1997).

\bibitem{OMD} M. C. de Oliveira, S. S. Mizrahi, and V. V. Dodonov, 
\emph{J. Opt. B: Quantum Semiclass. Opt.} {\bf 1}, 610 (1999).

\bibitem{CDM} A. S. M. de Castro, V. V. Dodonov, and S. S. Mizrahi, 
\emph{J. Opt. B: Quantum Semiclass. Opt.} {\bf 4}, S191 (2002).

\bibitem{Cel01}
E. Celeghini, M. Rasetti, and G. Vitiello, Phys. Rev. Lett. {\bf 66}, 2056 (1991).

\bibitem{Cel02}
E. Celeghini, M. Rasetti, and G. Vitiello, Annals of Physics {\bf 215}, 156 (1992).\

\bibitem{Cel03}
E. Celeghini, M. Rasetti, M. Tarlini and G. Vitiello, Mod. Phys. Lett. {\bf B 3}, 1213 (1989).

\bibitem{Vitiello}
G. Vitiello, Phys. Lett. {\bf A 376}, 2527 (2012). 

\bibitem{Vitiello4}
Y. M. Bunkov, H. Godfrin, \emph{Topological Defects and the 
Nonequilibrium Dynamics of Symmetry Breaking Phase Transitions}, 
Kluwer Academic Publ., Dordrecht, (2000).

\end{thebibliography}
\end{document}